\input harvmac
\input epsf.tex



\lref\WittenYC{
E.~Witten,
``Phases of ${\cal N} = 2$ theories in two dimensions,''
Nucl.\ Phys.\ B {\bf 403}, 159 (1993);
{\tt arXiv:hep-th/9301042}.
}

\lref\ColemanUZ{
S.~R.~Coleman,
``More about the massive Schwinger model,''
Annals Phys.\  {\bf 101}, 239 (1976).
}

\lref\first{
J.~Maldacena and H.~Ooguri,
``Strings in $AdS_3$ and $SL(2,R)$ WZW model. Part 1: the spectrum,''
J.\ Math.\ Phys. 42 (2001) 2929
({\sl Special issue on M Theory}); {\tt arXiv/0001053}.
}

\lref\second{J.~Maldacena, H.~Ooguri and J.~Son,
``Strings in $AdS_3$ and the $SL(2,R)$ WZW model.
Part 2: Euclidean black hole,''
J.\ Math.\ Phys.  42 (2001) 2961
({\sl Special issue on M Theory}); {\tt
arXiv/0005183}.
}

\lref\siw{E.~Silverstein and E.~Witten,
``Criteria for conformal invariance of $(0,2)$ models,''
Nucl.\ Phys.\ B {\bf 444}, 161 (1995);
{\tt arXiv:hep-th/9503212}.
}

\lref\wittenjones{E.~Witten,
``Chern-Simons gauge theory as a string theory,''
{\tt arXiv:hep-th/9207094}.
}

\lref\wittenknot{E. Witten, ``Quantum field theory
and the Jones polynomial,'' Comm. Math. Phys. {\bf 121}, 351
(1989).}

\lref\bcov{M.~Bershadsky, S.~Cecotti, H.~Ooguri and C.~Vafa,
``Kodaira-Spencer theory of gravity and exact
results for quantum string amplitudes,''
Commun.\ Math.\ Phys.\  {\bf 165}, 311 (1994);
{\tt arXiv:hep-th/9309140}.
}

\lref\agnt{
I.~Antoniadis, E.~Gava, K.~S.~Narain and T.~R.~Taylor,
``Superstring threshold corrections to Yukawa couplings,''
Nucl.\ Phys.\ B {\bf 407}, 706 (1993);
{\tt arXiv:hep-th/9212045}.
}

\lref\ovknot{
H.~Ooguri and C.~Vafa,
``Knot invariants and topological strings,''
Nucl.\ Phys.\ B {\bf 577}, 419 (2000),
{\tt arXiv:hep-th/9912123}.
}

\lref\gova{R.~Gopakumar and C.~Vafa,
``On the gauge theory/geometry correspondence,''
Adv.\ Theor.\ Math.\ Phys.\  {\bf 3}, 1415 (1999),
{\tt arXiv:hep-th/9811131}.
}

\lref\oova{
H.~Ooguri and C.~Vafa,
``Worldsheet derivation of a large $N$ duality,''
Nucl.\ Phys.\ B {\bf 641}, 3 (2002),
{\tt arXiv:hep-th/0205297}.
}

\lref\vaug{C.~Vafa,
``Superstrings and topological strings at large $N$,''
J.\ Math.\ Phys.\  {\bf 42}, 2798 (2001),
{\tt arXiv:hep-th/0008142}.
}

\lref\civ{F.~Cachazo, K.~Intriligator and C.~Vafa,
``A large $N$ duality via a geometric transition,''
Nucl. \ Phys. \ B {\bf 603}, 3 (2001),
{\tt arXiv:hep-th/0103067}.}

\lref\dv{
R.~Dijkgraaf and C.~Vafa,
``Matrix models, topological strings, and supersymmetric gauge theories,''
Nucl.\ Phys.\ B {\bf 644}, 3 (2002), {\tt
arXiv:hep-th/0206255};
``On geometry and matrix models,''
Nucl.\ Phys.\ B {\bf 644}, 21 (2002),
{\tt arXiv:hep-th/0207106};
``A perturbative window into non-perturbative physics,''
{\tt arXiv:hep-th/0208048};
``${\cal N}=1$ supersymmetry, deconstruction, and bosonic gauge theories,''
{\tt arXiv:hep-th/0302011}.}

\lref\gri{
R.~Dijkgraaf, M.~T.~Grisaru, C.~S.~Lam, C.~Vafa and D.~Zanon,
``Perturbative computation of glueball superpotentials,''
{\tt arXiv:hep-th/0211017}.
}

\lref\berktend{
N.~Berkovits,
``Super-Poincare covariant quantization of the superstring,''
JHEP {\bf 0004}, 018 (2000),
{\tt arXiv:hep-th/0001035}.
}

\lref\berkrr{
N.~Berkovits,
``Quantization of the superstring in Ramond-Ramond backgrounds,''
Class.\ Quant.\ Grav.\  {\bf 17}, 971 (2000),
{\tt arXiv:hep-th/9910251}.
}

\lref\bv{
N.~Berkovits and C.~Vafa,
``${\cal N}=4$ topological strings,''
Nucl.\ Phys.\ B {\bf 433}, 123 (1995),
{\tt arXiv:hep-th/9407190}.
}

\lref\berkcy{
N.~Berkovits,
``Covariant quantization of the Green-Schwarz superstring
in a Calabi-Yau background,''
Nucl.\ Phys.\ B {\bf 431}, 258 (1994),
{\tt arXiv:hep-th/9404162}.
}
\lref\CornalbaCU{
L.~Cornalba, M.~S.~Costa and R.~Schiappa,
``D-brane dynamics in constant Ramond-Ramond
potentials and  noncommutative geometry,''
{\tt arXiv:hep-th/0209164}.
}

\lref\vann{J.H. Schwarz and P. van Nieuwenhuizen, ``Speculations
concerning a fermionic structure of space-time,'' Lett. Nuovo Cim.
{\bf 34},21 (1982).}
\lref\ferr{S. Ferrara and M.A. Lledo, ``Some aspects of deformations
of supersymmetric field theories,'' JHEP {\bf 0005}, 008 (2000),
{\tt arXiv:hep-th/0002084}.}
%
\lref\KlemmYU{
D.~Klemm, S.~Penati and L.~Tamassia,
``Non(anti)commutative superspace,''
{\tt arXiv: hep-th/0104190}.
}
\lref\neww{J.~de Boer, P.~A.~ Grassi, P.~ van Nieuwenhuizen,
``Non-commutative superspace from string theory,''
{\tt arXiv:hep-th/0302078}.}

\lref\distler{
J.~Distler and C.~Vafa,
``A critical matrix model at $c = 1$,''
Mod.\ Phys.\ Lett.\ A {\bf 6}, 259 (1991).
}

\lref\penner{R.~C.~Penner, ``Perturbative series and the
moduli space of Riemann surfaces,'' J. Diff. Geom. {\bf 27}
(1988) 35.}
\lref\vy{
G.~Veneziano and S.~Yankielowicz,
``An effective Lagrangian for the pure ${\cal N}=1$
supersymmetric Yang-Mills theory,''
Phys.\ Lett.\ B {\bf 113}, 231 (1982).
}

\lref\govam{R.~Gopakumar and C.~Vafa,
``M-theory and topological strings. I,''
{\tt arXiv:hep-th/9809187};
``M-theory and topological strings. II,''
{\tt arXiv:hep-th/9812127}.
}
\lref\wittenbaryon{
E.~Witten,
``Baryons and branes in anti de Sitter space,''
JHEP {\bf 9807}, 006 (1998);
{\tt arXiv:hep-th/9805112}.
}

\lref\gross{
D.~J.~Gross and H.~Ooguri,
``Aspects of large $N$ gauge theory dynamics as seen by string theory,''
Phys.\ Rev.\ D {\bf 58}, 106002 (1998);
{\tt arXiv:hep-th/9805129}.
}

\lref\lastpaper{
H.~Ooguri and C.~Vafa,
``The $C$-deformation of gluino and non-planar diagrams,''
{\tt arXiv:hep-th/0302109}.
}


{\Title{\vbox{
\hbox{CALT-68-2433}
\hbox{HUTP-03/A020}
\hbox{\tt hep-th/0303063}}}
{\vbox{
\centerline{Gravity Induced $C$-Deformation}}}
\vskip .3in
\centerline{Hirosi Ooguri$^1$ and Cumrun Vafa$^{1,2}$}
\vskip .4in
}

\centerline{$^1$ California Institute of Technology 452-48,
Pasadena, CA 91125, USA}
\vskip .1in
\centerline{$^2$ Jefferson Physical Laboratory, Harvard University,
Cambridge, MA 02138, USA}

\vskip .4in

We study F-terms describing coupling of the supergravity
to ${\cal N}=1$ supersymmetric gauge theories which admit
large $N$ expansions. We show that these F-terms are given
by summing over genus one non-planar diagrams of the large
$N$ expansion of the associated matrix model (or more generally
bosonic gauge theory).  
The key ingredient in this derivation is the observation that
 the chiral ring of the gluino
fields is deformed by the supergravity
fields, generalizing
the $C$-deformation which was recently introduced. The gravity
induced part of the $C$-deformation can be derived from
the Bianchi identities of the supergravity, but understanding
gravitational corrections to the F-terms requires
a non-traditional interpretation of these identities.

\vfill
\eject

\newsec{Introduction}

The connection between supersymmetric
gauge theories and matrix models (or more generally
bosonic gauge theories) has led to exact non-perturbative
computation of F-terms starting from perturbative
computations in the gauge theory \dv .  In the context of gauge
theory on flat space, only the planar diagrams are relevant
for the computation of F-terms.  However if one
goes beyond flat space or consider certain
deformations, it is expected that the non-planar
diagrams become relevant for computing F-terms.
In particular in a recent paper \lastpaper ,
we introduced the notion of the $C$-deformation of
${\cal N}=1$ gauge theories. Without the deformation, the gluino
fields ${\cal W}_\alpha$ in these theories satisfy the chiral ring relation,
\eqn\ring{ \{ {\cal W}_\alpha,  {\cal W}_\beta \} = 0,}
as pointed out in \ref\aharony{O.~Aharony, {\it unpublished},
as quoted in,
C.~Vafa,
``Puzzles at large N,''
{\tt arXiv:hep-th/9804172}.
}.
This relation
plays an important role in classifying chiral primary fields
in these theories. In \lastpaper , we showed that a self-dual
two-form $F_{\alpha\beta}$ can be used to deform this relation
as
\eqn\cdeform{ \{ {\cal W}_\alpha, {\cal W}_\beta \} = F_{\alpha \beta}.}
In string theory, $F_{\alpha\beta}$ has the interpretation as
the graviphoton field strength of the ${\cal N}=2$ supergravity
coupled to the branes. 
We can view this as the {\it defining property} of the
gluino fields, modifying the condition that they be
Grassmannian variables.
We called this the $C$-deformation and showed
that the non-planar diagrams of matrix models
 captures the $F_{\alpha \beta}F^{\alpha
\beta}$ dependence of the glueball superpotential.

Another place where non-planar diagrams should enter involves
gravitational corrections. In particular it was conjectured
in \dv\ that certain $R^2$ type terms can be computed
exactly by studying the non-planar perturbative gauge theory amplitudes
with a single handle. They are
expressed in terms of the glueball fields and
evaluated at the extremum of
the superpotential computed by the planar diagrams.  This conjecture
was motivated by the meaning of topological string amplitudes
in the context of low energy effective theories of superstring
compactifications
\refs{\bcov,\agnt,\ovknot,\vaug}\ together with the large
$N$ duality conjectures \gova , proven in \oova\ and embedded
in superstrings in \vaug .  This prediction
has already been tested in a number of cases: for the
gravitational correction for ${\cal N}=4$ Yang-Mills in
the third paper in \dv , and for certain ${\cal N}=2$ supersymmetric
gauge systems in \lref\mar{A. Klemm, M. Marino,
 S. Theisen, ``Gravitational corrections in supersymmetric gauge
theory and matrix models,'' {\tt arXiv:hep-th/0211216.
}}\lref\dijet{R. Dijkgraaf, A. Sinkovics, M. Temurhan,
``Matrix models and gravitational corrections,''
{\tt arXiv:hep-th/0211241.}}
\refs{\mar,\dijet}. Our aim in this paper is to prove this conjecture.

 In \lastpaper , we showed
that the effective superpotential of the $C$-deformed gauge theory
\cdeform\
is computed by the full matrix model partition function
including non-planar diagrams. More explicitly, if we define
the glueball superfield $S_i$ by
\eqn\glueball{ S_i = {1\over 32\pi^2} \epsilon^{\alpha\beta}
              {\rm Tr}_i {\cal W}_\alpha {\cal W}_\beta,}
where ${\rm Tr}_i$ is over the $i$-th gauge group of rank $N_i$,
their effective superpotential is given by
\eqn\cdeformedpotential{
\Gamma_1 = \sum_{g=0}^\infty~\int d^4 x d^2\theta
\ (F_{\alpha\beta} F^{\alpha\beta})^g \ N_i {\partial F_g\over
\partial S_i} (S),}
where $F_g$ is given by the matrix model partition function computed
by a sum over genus $g$ diagrams with $S_i$ playing
the role of the 't Hooft loop counting parameters.
  There is another series
of gravitational corrections predicted in \refs{\bcov,\vaug}, which
takes the form,
\eqn\gravicorrections{
\Gamma_2 =\sum_{g=1}^\infty ~g \int d^4 x d^2\theta ~
{\cal W}_{\alpha\beta\gamma} {\cal W}^{\alpha\beta\gamma}
(F_{\rho\sigma}F^{\rho\sigma})^{g-1} F_g(S),}
where ${\cal W}_{\alpha\beta\gamma}$ denotes the
${\cal N}=1$ gravitino superfield.
In this paper, we will show that \cdeformedpotential\ continues to hold
and
\gravicorrections\ computes
the mixed gravitational/glueball superpotential of the
 ${\cal N}=1$ gauge theory
if we postulate that the gluino fields obey the relation
\eqn\deformation{
  \{ {\cal W}_\alpha ,  {\cal W}_\beta \} =2
{\cal W}_{\alpha\beta\gamma}
{\cal W}^\gamma+F_{\alpha\beta}.}
If we set the Lorentz violating parameter $F_{\alpha \beta}=0$,
only planar contribution in \cdeformedpotential\ and genus one contribution
in \gravicorrections\ survive.
{}From the string theory point of view, the relation arises
as follows. The supersymmetry
variation of the open string worldsheet with the
gravitino background ${\cal W}_{\alpha\beta\gamma}$ gives
rise to boundary terms. We can cancel these boundary terms
and restore the supersymmetry if we assume this relation
\deformation . This is essentially the same as the way
we derived the $C$-deformation \cdeform\ for the
graviphoton background.

It turns out that, when $F_{\alpha\beta}=0$, the relation
\deformation\ can also be understood in the conventional framework
of supergravity theory --- it follows from the supergravity
tensor calculus. This is in contrast to the
deformation by $F_{\alpha\beta}$, which does not have such
a conventional interpretation via ${\cal N}=1$ supersymmetry.
However, we will point out that a proper interpretation of the
gravitational corrections \gravicorrections\ requires
a non-traditional interpretation of this standard
relation.  In particular in the case of $U(1)$ gauge theories,
the traditional interpretation of \deformation\ would be that
the left-hand side and the right-hand side of the equations vanish
separately (the left-hand side being zero is due to the
Grassmannian property of ${\cal W}_a$, and this forces
the right-hand side to be equal to zero also).  However
we shall find that preservation of supersymmetry in the presence
of constant ${\cal W}_{\alpha \beta \gamma}$ gravitational
background requires only the weaker relation where
we postulate \deformation\ but do not impose
the standard Grassmannian
properties on ${\cal W}_\alpha$.  Despite this non-traditional
interpretation,  this seems to be the natural choice
since supersymmetry only requires the weaker relation
and it is the one that leads to large
$N$ dualities in superstring theory. In particular without
this non-traditional interpretation of the relation
\gravicorrections , we shall see that the large $N$ superstring
duality proposed in \vaug\ would not hold.

It turns out that there are also planar contributions to
superpotential terms of the form ${\cal W}_{\alpha \beta \gamma}
{\cal W}^{\alpha \beta \gamma} \ S^n$.
However as will be shown in  \ref\newp{
R. Dijkgraaf, M. Grisaru, H. Ooguri, C. Vafa and
D. Zanon, {\it in preparation}.}, once one substitutes the
expectation value for the glueball field which extremizes
 the superpotential, this
contribution becomes trivial. This is also consistent with the large $N$
superstring duality \vaug\
since there is no $R^2$ correction coming from genus $0$
on the closed string dual.

This paper is organized as follows. In section 2, we derive
the relation \deformation\ from the point of view of string
theory. In section 3, we derive the same relation from the
supergravity tensor calculus. In section 4, we show that this
deformation leads to the gravitational corrections \gravicorrections\
including the more general situations not necessarily embedded in string
 theory.  We also discuss certain mixed gravitational/gauge interactions
which violate Lorentz invariance
and which could serve as an experimental signature for the $C$-deformation.

\newsec{Deformation of the chiral ring I: string theory perspective}

In this section, we consider gravitational corrections to the ${\cal N}=1$
gauge theory in four dimensions which is defined as the low energy limit
of Type II superstring with D($N+3$) branes wrapping on $n$ cycles on
a Calabi-Yau three-fold and extending in four flat dimensions.  We will
concentrate on the universal spacetime part of this computation.
Even though the string context may appear to be restrictive (in that
one is limited to field theories arising from string theory),
the more general field theory setup discussed in \gri\
can be effectively related to the spacetime part of the string
computation, as we have demonstrated in our previous paper
\lastpaper . In the string context the perturbative computation
is better organized since one worldsheet topology corresponds
to many Feynman diagrams. Once we understand what is going on in
string theory, we can directly translate each step
into the more general field theory context. This
is the reason why we start our discussion from the string theory
perspective.

The F-terms of the low energy effective
theory are given by \cdeformedpotential\ and \gravicorrections , where
\eqn\whatfg{ F_g(S) =\sum_{h=0}^\infty F_{g,h} S^h,}
and $F_{g,h}$ is the topological string partition function
for genus $g$ worldsheet with $h$ boundaries ending on
D branes wrapping on these cycles.\foot{For simplicity,
we consider the case with a single cycle. Correspondingly,
there is only one boundary-counting parameter $S$. Generalization
to cases with more cycles is straightforward.} According
to \wittenjones , these
topological string partition functions can be computed using
the Chern-Simons theory (or its dimensional reduction). In
particular, for a specific class of D5 branes wrapping on 2-cycles,
the dimensional reduction of the Chern-Simons theory turns out
to be a matrix model \dv .

In the previous paper \lastpaper , we explained how the gravitational
corrections of the type \cdeformedpotential\ arises from the
string theory computation and showed that it can also be obtained
from purely gauge theoretical Feynman diagram computation if we
deform the chiral ring as \cdeform . In this paper, we study
the second series of gravitational corrections \gravicorrections .
As in the previous paper, we start our discussion on the string
worldsheet, which we describe using the covariant quantization
of superstring developed in \berkcy . As demonstrated in \bv ,
this is the most economical way to establish the relation between
topological string amplitudes and the F-terms in Type II superstring
compactified on a Calabi-Yau three-fold, which was originally
derived in the NSR formalism in \refs{\bcov,\agnt}.  In the formalism
of \berkcy , the
four-dimensional part of the worldsheet Lagrangian density
that is relevant for our discussion is simply given by
\eqn\covariant{
 {\cal L} = {1\over 2} \partial X^\mu \bar \partial X_\mu
  + p_\alpha \bar \partial \theta^\alpha +
p_{\dot \alpha} \bar \partial \theta^{\dot \alpha}+
 \bar{p}_\alpha
\partial \bar \theta^\alpha + \bar p_{\dot \alpha}
\partial \bar \theta^{\dot \alpha},}
where $p$'s are $(1,0)$-forms, $\bar p$'s are
$(0,1)$-forms, and $\theta ,\bar \theta$'s are $0$-forms.
The remainder of the Lagrangian density consists of the topologically
twisted ${\cal N}=2$
supersymmetric sigma-model on the Calabi-Yau three-fold and
a chiral boson which is needed to construct the R current.
We work in the chiral representation of supersymmetry, in
which spacetime supercharges are given by
\eqn\supercharges{ \eqalign{
&Q_\alpha = \oint p_\alpha \cr
&Q_{\dot \alpha} = \oint p_{\dot \alpha} - 2i\theta^\alpha
 \partial X_{\alpha \dot \alpha}
 + \cdots,}}
where $X_{\alpha\dot \alpha} = \sigma_{\alpha\dot\alpha}^\mu X_\mu$,
and $\cdots$ in the second line represents terms
containing $\theta^{\dot \alpha}$ and
$\theta^2 = \epsilon_{\alpha\beta}\theta^\alpha \theta^\beta$.
The second set of supercharges $\bar Q_\alpha, \bar Q_{\dot \alpha}$
are defined by replacing $p, \theta$ by
$\bar p, \bar\theta$. These generate the ${\cal N}=2$ supersymmetry
 in the bulk.
When the worldsheet is ending on D branes and extending in four
dimensions, the boundary conditions for the worldsheet variables
are given by
\eqn\boundarycondition{\eqalign{
  & (\partial - \bar\partial) X^\mu = 0, \cr
  & \theta^\alpha = \bar \theta^\alpha, ~~p_\alpha = \bar p_\alpha.}}
Here we assume that the boundary is located at ${\rm Im}\ z=0$.
These boundary conditions preserve one half of the supersymmetry
generated by $Q+ \bar Q$.

In these conventions, the vertex operators for the graviphoton
$F_{\alpha\beta}$
and the gravitino ${\cal W}_{\alpha\beta\gamma}$ are given by
\eqn\graviphotonvertex{ \int F^{\alpha\beta} p_\alpha \bar p_\beta, }
and
\eqn\gravitinovertex{ \int {\cal W}^{\alpha\beta\gamma} \left(
 p_\alpha X_{\beta\dot\beta} \bar \partial X_{\gamma\dot\gamma}
+ \bar p_\alpha X_{\beta\dot\beta} \partial X_{\gamma\dot\gamma}
\right) \epsilon^{\dot\beta\dot\gamma}+
\int {\cal W}^{\alpha\beta\gamma}
p_\alpha \bar p_\beta (\theta_\gamma - \bar\theta_\gamma),}
respectively.
The gluino ${\cal W}_\alpha$ couples to the boundary $\gamma_i$
of the worldsheet ($i=1,\cdots,h$) as
\eqn\gluinovertex{ \oint_{\gamma_i} {\cal W}^\alpha p_\alpha.}

We can make a simple counting of fermion zero modes to determine
topology of worldsheets that contribute to a particular F-term.
On a genus $g$ surface with $h$ boundaries, there are $(2g+h-1)$
zero modes for each $p_\alpha$ ($\alpha=1,2$). One possible
ways to absorb these zero modes, as was done in \lastpaper , is to
 insert $2g$ graviphotons
and $2h-2$ gluinos. In order for these insertions to actually
absorb the zero modes, we need two gluinos for each boundary
except for one. We cannot insert gluinos on all boundaries
since the sum $\sum_{i=1}^h \gamma_i$ is homologically
trivial and
\eqn\sumvanishes{ \sum_{i=1}^h \oint_{\gamma_i}p_\alpha=0.}
Therefore the topological string computation on
genus $g$ worldsheet with $h$ boundaries gives
the combination $N h S^{h-1} (F^2)^g$,
where the factor $N$ comes from the gauge group trace
on the boundary where the gluino is not inserted,
$h$ comes from the choice of such a boundary, each
boundary with gluino insertion is counted with the factor
$S = {\rm Tr} \ {\cal W}^2$, and we have $2g$ graviphoton
insertions. As we pointed out in \lastpaper , there
is more to the story --- in order to correctly reproduce
the F-term computation, we need to take into account
the effect due to the deformation of the chiral ring
\cdeform\ --- but the counting of the zero modes
is correct as it is. Taking into account the $C$-deformation,
we found in the previous paper that
the F-term contribution from genus $g$ worldsheet with
$h$ boundaries is $F_{g,h}$, and it
can be expressed as a sum over the matrix model 't Hooft
diagrams of the corresponding topology.
This gives rise to the first series
of gravitational corrections \cdeformedpotential .

To understand the second series \gravicorrections ,
we need to consider two insertions of the gravitino
vertex operator \gravitinovertex .   For simplicity of discussion,
let us first turn off $F_{\alpha \beta}=0$.
There are two possible terms for gravitino vertex
operator.  Either one uses the first part of the gravitino
vertex operator \gravitinovertex\ which involves only one
$p$ or the second term which involves two $p$'s. We cannot
use mixed types, because that will not lead to absorption of all $p$ zero
modes.  Note that both types of terms have one net $p$ charge.
Thus we can absorb two net $p$ zero modes from the two gravitino
insertions ${\cal W}_{\alpha\beta\gamma}{\cal W}^{\alpha\beta\gamma}$.
To absorb
the rest, we will use the gluino fields on the boundary.  If we choose
$n$ boundaries and put two gluinos ${\cal W}_\alpha {\cal W}^\alpha$
on each, we have for the condition of the absorption
of the $p$ zero modes that
$$2n+2=2(2g+h-1) ~~~ \rightarrow ~~~ n=2g+h-2.$$
Since $n\leq h$, we have either $g=0$ and $n=h-2$ or
$g=1$ and  $n=h$. Namely possible F-terms are
${\cal W}_{\alpha\beta\gamma}{\cal W}^{\alpha\beta\gamma}
\ S^{h-2}$ from $g=0$ and
${\cal W}_{\alpha\beta\gamma}{\cal W}^{\alpha\beta\gamma} \ S^h$ from
$g=1$.

If we use the first term of the gravitino
vertex \gravitinovertex , we do not have an option of $g=1$
and $n=h$
since the gravitino vertex
 anti-commutes with $\oint_{\gamma_i} p_\alpha$ and
therefore  $\sum_{i=1}^h \oint_{\gamma_i} p_\alpha =0$.
Namely $\oint_{\gamma_i}p_\alpha$ are not linearly independent
and we cannot insert the gluino vertex operators on all
boundaries.   Thus it only contributes to
$g=0$ and $n=h-2$,
namely to {\it planar diagrams}.  These planar contributions will
be discussed in \newp , where it will be shown to be
 non-vanishing.  However, it will also be shown there that
their contributions to the F-terms
become trivial when we substitute the extremum value
of $S$, thus the planar contributions effectively
drop out, consistently
with the superstring duality in \vaug .

If we use the second part of
 the gravitino vertex operator
instead, the $g=1$ contribution does not vanish.
This is because the second term contains
$(\theta^\gamma-\bar \theta^\gamma)$, and it has
nontrivial correlation with $\oint p_\alpha$
on the boundary. The sum $\sum_i \oint_{\gamma_i} p_\alpha$
does not have to vanish, and we can insert gluino vertex
operators on all boundaries. In fact, a simple application
of the Cauchy integral formula gives
\eqn\sumdoesnotvanish{\sum_{i=1}^h
\oint_{\gamma_i} {\cal W}^\alpha p_\alpha \cdot
\int {\cal W}^{\alpha\beta\gamma}
p_\alpha \bar p_\beta (\theta_\gamma - \bar\theta_\gamma)
\sim
\int {\cal W}^{\alpha\beta\gamma}{\cal W}_\gamma~
p_\alpha \bar p_\beta.}
This can lead to non-zero result for $g=1$ and $n=h$,
giving rise to the gravitational correction of the
form ${\cal W}_{\alpha\beta\gamma}{\cal W}^{\alpha\beta\gamma}
\ S^h$ in \gravicorrections . As in our previous paper
\lastpaper , there is more to the story. The presence of the
gravitino background modifies the chiral ring of the
gluino field as
\eqn\gravideformation{
 \{ {\cal W}_\alpha, {\cal W}_\beta \} = 2{\cal W}_{\alpha\beta\gamma}
{\cal W}^\gamma. }
Taking this into account, we can reproduce the topological
string amplitude
$g \ F_{g,h}$ that multiplies  to ${\cal W}_{\alpha\beta\gamma}
{\cal W}^{\alpha\beta\gamma} \ S^h$ in \gravicorrections .
On the other hand, this effect does not give
contributions to planar diagrams. This is evident from the
presence of the factor $g$ in $g\ F_{g,h}$.

Let us explain how the deformation \gravideformation\ arises from the
string theory perspective. We
 follow the approach of \lastpaper\ and  look
at the variation of the gravitino vertex operator under
$\epsilon^{\dot \alpha}(Q+\bar Q)_{\dot\alpha}$. We find
\eqn\susyvariation{
 \delta\left[ \int {\cal W}^{\alpha\beta\gamma}
p_\alpha \bar p_\beta (\theta_\gamma - \bar \theta_\gamma)\right]
=
2i\epsilon^{\dot \alpha} {\cal W}^{\alpha\beta\gamma}
\int d\Big( Y_{\alpha\dot \alpha} (p_\beta+\bar p_\beta)
(\theta_\gamma - \bar \theta_\gamma)\Big) ,}
where
\eqn\whaty{ Y_{\alpha\dot \alpha} = X_{\alpha\dot\alpha}
+i \theta_\alpha \theta_{\dot \alpha}
+ i \bar \theta_{\alpha} \bar \theta_{\dot \alpha}.}
Since the integrand of the right-hand side of \susyvariation\
is total derivative and $\theta_\gamma=\bar\theta_\gamma$ on the boundaries,
it would vanish if there are no other operators inserted
on the boundaries. The only non-zero contribution comes from
the operator product singularity of \susyvariation\ with
the gluino vertex operator as
\eqn\nonzerosusyvariation{
\delta\left[ \int {\cal W}^{\alpha\beta\gamma}
p_\alpha \bar p_\beta (\theta_\gamma - \bar \theta_\gamma)\right]
\cdot \oint {\cal W}^\alpha p_\alpha
= 4\epsilon^{\dot \alpha} \oint {\cal W}^{\alpha\beta\gamma}{\cal W}_\gamma
\  Y_{\alpha\dot\alpha} p_\beta.}
Comparing with our previous paper (see eq. (2.21) of
\lastpaper\ and the subsequent discussion), we find that the
boundary terms can be cancelled by imposing the relation
\gravideformation . It is evident from \lastpaper\ that,
if the graviphoton $F_{\alpha\beta}$ is turned on,
this is further deformed as
\eqn\deformation{ \{ {\cal W}_\alpha, {\cal W}_\beta \}
 = 2{\cal W}_{\alpha\beta\gamma} {\cal W}^\gamma + F_{\alpha\beta}
~~{\rm mod}~\bar D . }
Note that the identity is modulo
$D_{\dot\alpha}$ since that is all we need to cancel the boundary
terms.

In the flat supergravity background,
the definition of the gluino superfield
\eqn\gluino{ {\cal W}_\alpha = {1\over 4i}[ D^{\dot \alpha},
 D_{\alpha\dot \alpha}]}
and the fact that this superfield is chiral $D_{\dot \alpha} {\cal W}_\beta=0$
imply \aharony ,
\eqn\undeformed{ \{ {\cal W}_\alpha, {\cal W}_\beta \} =
0 ~~~{\rm mod}~\bar D.}
As shown in \gri\ and
\ref\CachazoRY{
F.~Cachazo, M.~R.~Douglas, N.~Seiberg and E.~Witten,
``Chiral rings and anomalies in supersymmetric gauge theory,''
JHEP {\bf 0212}, 071 (2002),
{\tt arXisv:hep-th/0211170}.
} using direct field theory analysis, the effective superpotential
in this case receives contributions only from planar diagrams,
consistently with the topological string computation discussed in the above.
In section 4, we will show that the superpotential for
the gluino obeying the deformed relation \deformation\ is computed by
the full partition function of the matrix model
including non-planar diagrams
and reproduce the gravitational corrections \gravicorrections\ as well
as \cdeformedpotential\ predicted
by the topological string computation \bcov\ and the
large $N$ duality \vaug .

\newsec{Deformation of the chiral ring II: supergravity perspective}

It turns out that the gravitino part of the deformed
chiral ring relation \deformation\
\eqn\gravideform{ \{ {\cal W}_\alpha, {\cal W}_\beta\}
= 2{\cal W}_{\alpha\beta\gamma}{\cal W}^\gamma,}
can also be understood from the standard supergravity tensor calculus
\ref\GatesNR{See, for example,
S.~J.~Gates, M.~T.~Grisaru, M.~Rocek and W.~Siegel,
``Superspace, or one thousand and one lessons in supersymmetry,''
Front.\ Phys.\  {\bf 58}, 1 (1983),
{\tt arXiv:hep-th/0108200}.
}.
In fact, the Bianchi identity implies\foot{We are ignoring the
other chiral superfield $R$ which appears as the torsion in
$[D_\alpha, D_{\beta\dot\beta}]$ since it vanishes on-shell.}
\eqn\bianchiid{
[ D^{\dot \alpha}, D_{\alpha\dot\alpha}]_{\beta\gamma}
= 4i {\cal W}_\alpha\epsilon_{\beta\gamma}
-8i {\cal W}_{\alpha\beta\gamma},}
where we are considering these operators acting
on chiral spinor superfields (which is why we have 
spinor indices $\beta \gamma$ in the above).  The second 
term above arises from the Lorentz action on the spinor field.
Let us repeat the derivation of \undeformed\ in
the supergravity background using this relation. We use the fact that
${\cal W}_{\dot \alpha}$ is chiral to show
$$\left\{ [ D^{\dot \alpha}, D_{\alpha\dot \alpha}], {\cal W}_\beta
\right\} =
\left\{ D^{\dot \alpha}, \left[ D_{\alpha\dot \alpha}, {\cal W}_\beta
\right] \right\}
= 0 ~~~{\rm mod}~\bar D.$$
Substituting \bianchiid\ to the left-hand side of this equation,
we find
\eqn\sugraderivation{
\{ {\cal W}_\alpha, {\cal W}_\beta\}
-2{\cal W}_{\alpha\beta\gamma} {\cal W}^\gamma
= 0.}
again modulo $\bar D$. This is what we wanted to show.
We have found that the gravitino part of the deformation \deformation\
is due to the standard supergravity tensor calculus.
However
a proper understanding of the F-terms \gravicorrections\
requires a non-traditional interpretation of this relation,
as we shall see below.

\newsec{Non-planar diagrams in the field theory limit}

The field theory limit of the above string theory
computation is straightforward, and is very similar
to our discussion in the
previous paper \lastpaper. We will only
point out some salient features.
In \lastpaper , the graviphoton vertex operator
$\int F^{\alpha \beta}p_{\alpha}\bar p_{\beta}$
disappears in the field theory limit, where $p=\bar p$.
Its effect, however, survives if we include the $C$-deformation
on the gluino fields. Similarly here, the relevant
part of the gravitino vertex operator, the second term
in \gravitinovertex ,
vanishes in the field theory limit.  Effects of the gravitino
background survives
in the field theory if we include the $C$-deformation
for the gluino field, which as we discussed before would be needed
if we wish to preserve supersymmetry and in fact follows from
the supergravity tensor calculus.  Note that here
we still have a choice on algebraic properties of the fields, and it is
not dictated just from the tensor calculus leading to \gravideform .
For example consider the case where we consider the ${\cal W}_1$
component of a $U(1)$ gauge field
and suppose ${\cal W}_{111}$ background is non-zero.
 Then the chiral relation \gravideform\ gives
$$({\cal W}_1)^2- {\cal W}_{111}{\cal W}_2=0 ~~~{\rm mod}~\bar D$$
So far, this is perfectly standard supergravity tensor calculus
as discussed in the last section. However, what does one take
$({\cal W}_1)^2$ to be?  Usually
we set it to zero by the Grassmannian property of the gluino field
${\cal W}_\alpha$, which would then mean that
${\cal W}_{111} {\cal W}_2=0$ modulo $\bar D$.
It would just mean that this term is not going to appear
in any F-term, as it
is trivial as a chiral superfield. Thus we would have found
no corrections involving mixed gravitational/glueball fields
for non-planar diagrams in contradiction with the large $N$ duality
\vaug .
 This is what one would obtain in the standard, non
$C$-deformed treatment of Feynman diagrams.  However the $C$-deformation
we consider is the weaker statement which requires \gravideform\
but does not postulate the additional condition that
the gluino fields are Grassmannian variables.
In particular we {\it do not} require $({\cal W}_1)^2=0$.
This is how we end up getting a non-trivial result
from Feynman diagrams, following the discussion in \lastpaper .

Given the relation \deformation , it is straightforward
to reproduce the two series of gravitational corrections
\cdeformedpotential\ and \gravicorrections\ from purely
field theory Feynman diagram computations. In our previous
paper \lastpaper , we have shown how this is done for
the first series \cdeformedpotential\ when we have
the $C$-deformation,
\eqn\cdeformagain{ \{ {\cal W}_\alpha, {\cal W}_\beta \} =
F_{\alpha\beta}.}
By simply replacing $F_{\alpha\beta}$ by
$F_{\alpha\beta} + 2{\cal W}_{\alpha\beta\gamma}{\cal W}^\gamma$
and noting,
\eqn\traceformula{\eqalign{&
{\rm Tr}\Big[ (F_{\alpha\beta}+2{\cal W}_{\alpha\beta\gamma}
{\cal W}^\gamma)(F^{\alpha\beta} +2 {\cal W}^{\alpha\beta}_{~~\gamma}
{\cal W}^\gamma) \Big]^g \cr
&= N (F_{\alpha\beta} F^{\alpha\beta})^g
 + 4g (F_{\alpha\beta} F^{\alpha\beta})^{g-1}
\ ({\cal W}_{\alpha\beta\gamma}{\cal W}^{\alpha\beta\gamma})
\ {\rm Tr}\ {\cal W}_\alpha {\cal W}^\alpha,}}
we see that the deformation \deformation\ generates
both types of the F-terms simultaneously.
%

\subsec{Other corrections}

Note that the full corrections expected from string
theory  (including
the $U(1)$ fields ${\cal W}_\alpha$)
to the F-terms can be summarized by
the term \vaug ,
$$\Gamma =\int d^4x d^2\theta \int d^2{\hat \theta}
\Big[(F_{\alpha \beta}+\hat \theta^\gamma {\cal W}_{\alpha
\beta \gamma})(F^{\alpha \beta}+\hat \theta_\delta {\cal W}^{\alpha
\beta \delta })\Big]^g F_g(S+\hat \theta^{\alpha}{\cal W}_{\alpha} +\hat
\theta^2 N) .$$
This includes, in addition to the terms already discussed in this paper
and the previous paper \lastpaper ,  some mixed terms
involving the $U(1)$ superfield ${\cal W}_{\alpha}$ and the
gravitino superfield of the form,
$$2g \int d^4x d^2\theta F^{\alpha \beta}
{\cal W}_{\alpha \beta \gamma} {\cal W}^{\gamma}
\cdot (F^2)^{g-1} {\partial F_g \over \partial S} .$$
These terms can also derived easily along the lines
we have discussed here and in the previous paper.  There
is one very interesting aspect of these terms, however, that
we wish to point out. In the background of non-zero
graviphoton field strength, these terms generate
``photon/graviton interactions'' which violate Lorentz
invariance.  As noted in \lastpaper ,
the non-gravitational F-terms have the property that they
screen violation of Lorentz invariance in the graviphoton
background. The terms we are finding here
after integration over the $d^2\theta$ will involve
terms like $F^{U(1)} R$ where
$F^{U(1)}$ is the field strength in the gluino multiplet
and the indices are contracted appropriately with 
the Lorentz violating parameter $F_{\alpha \beta}$.
If $F_{\alpha\beta} \neq 0$, this generates
non-Lorentz invariant mixing of the photon and the graviton.
They could give interesting
signatures of the $C$-deformation, and it would be amusing
to see  if it is realized in Nature.

\bigskip
\bigskip

\centerline{{\bf Acknowledgments }}

\bigskip
We thank R. Dijkgraaf, M. Grisaru, and D. Zanon, whose
insights have contributed significantly to this
paper.  We are
also grateful to N. Berkovits for valuable discussion.

C.V. thanks the hospitality of the
theory group at Caltech, where he is a Gordon Moore
Distinguished Scholar.

The research of H.O. was supported in part by
DOE grant DE-FG03-92-ER40701.  The research of C.V. was supported
in part by NSF grants PHY-9802709 and DMS-0074329.

\listrefs

\end